\documentclass[%
 reprint,
superscriptaddress,
 amsmath,amssymb,
 aps,
]{revtex4-2}
\usepackage{soul}
\usepackage{rotating}
\usepackage{graphicx}
\usepackage{dcolumn}
\usepackage{bm}
\usepackage[dvipsnames]{xcolor}
\usepackage{hyperref}
\usepackage{mathptmx}
\usepackage{ulem}
\hypersetup{
    colorlinks=true,
    citecolor=blue,
    linkcolor=blue,
    urlcolor=blue
    }

\begin{document}

\preprint{APS/123-QED}


\title{Ultrafast antiferromagnetic switching of Mn$_2$Au with laser-induced optical torques}

\author{Jackson L. Ross}
\affiliation{School of Physics, Engineering and Technology, University of York, YO10 5DD, York, United Kingdom}
\author{Paul-Iulian Gavriloaea}
\affiliation{Instituto de Ciencia de Materiales de Madrid, CSIC, Cantoblanco, 28049 Madrid, Spain}
\author{Frank Freimuth}
\affiliation{Institute of Physics, Johannes Gutenberg University Mainz, 55099 Mainz, Germany}
\affiliation{Peter Grünberg Institut and Institute for Advanced Simulation, Forschungszentrum Jülich and JARA, 52425 Jülich, Germany}
\author{Theodoris Adamantopoulos}\affiliation{Peter Grünberg Institut and Institute for Advanced Simulation, Forschungszentrum Jülich and JARA, 52425 Jülich, Germany}
\author{Yuriy Mokrousov}
\affiliation{Peter Grünberg Institut and Institute for Advanced Simulation, Forschungszentrum Jülich and JARA, 52425 Jülich, Germany}
\affiliation{Institute of Physics, Johannes Gutenberg University Mainz, 55099 Mainz, Germany}
\author{Richard F.L. Evans}
\affiliation{Department of Physics, University of York, YO10 5DD, York, United Kingdom}
\author{Roy Chantrell}
\affiliation{Department of Physics, University of York, YO10 5DD, York, United Kingdom}
\author{Rubén M. Otxoa}
\affiliation{Hitachi Cambridge Laboratory, CB3 OHE Cambridge, United Kingdom}
\affiliation{Donostia International Physics Center, 20018 Donostia San Sebastian, Spain}
\author{Oksana Chubykalo-Fesenko}
\affiliation{Instituto de Ciencia de Materiales de Madrid, CSIC, Cantoblanco, 28049 Madrid, Spain}

\date{\today}

\begin{abstract}
Efficient manipulation of the N\'eel vector in antiferromagnets can be  induced by generation of spin-orbit (SOT) or spin-transfer (STT) torques. Here we predict another possibility for antiferromagnetic domain switching using a non-zero staggered field induced from optical laser excitation. We present results from atomistic spin dynamics simulations from the application of a laser-induced torque using optical frequencies for all-optical switching (AOS) of the N\'eel vector in the antiferromagnet Mn$_2$Au. The driving mechanism takes advantage of the sizeable exchange enhancement characteristic of antiferromagnets, allowing for picosecond 90 and 180 degree precessional toggle switching with laser fluences on the order of mJ/cm$^2$. The symmetry of these novel torques is highly dependent on the time-varying  magnetisation direction, creating a sign change in the torque which greatly minimises the over-shooting effect common to SOT and STT. Lastly, we demonstrate the opportunity for this laser optical torque to produce deterministic, non-toggle switching of single antiferromagnetic domains.
\end{abstract}
\maketitle

\section{\label{sec:intro} Introduction}

Classes of high-N\'eel temperature antiferromagnets (AFMs), which respond asymmetrically to external stimuli, represent an important step for Terahertz (THz) radiation generation, as well as for manipulation and characterisation of magnetic textures \cite{Zelezny2017}. Currently, the most promising method for controlling the AFM order parameter remains the use of staggered spin orbit fields to generate a non-staggered spin orbit torque (SOT) which will drive the N\'eel vector precessionally~\cite{Bodnar2018,Roy2016,Selzer2022,Chen2019}. Also of interest is the use of spin transfer torques (STTs) for exciting AFM switching \cite{Cheng_2015,Weienhofer2023,Huang2022}. However, limitations to these methods provide challenges for practical implementation. SOT control using applied currents requires either precise timing to prevent over-switching \cite{Roy2016,Cheng_2015,Gomonay_2010}, weaker fields for longer durations \cite{RamaEiroa2022}, or repeated short pulses \cite{Bodnar2018,Olejn_k_2018}. STT driven switching requires complex heterostructures \cite{Weienhofer2023,Gomonay_2010}, or strong laser pulses \cite{Weienhofer2023}, and overswitching beyond 90 degrees is still a risk \cite{Cheng_2015,Weienhofer2023}. With Mn$_2$Au and CuMnAs remaining the only two materials with experimental confirmation of staggered spin-orbit torques produced by external current \cite{Wadley2016,Bodnar2018,godinho2018electrically,janda2020magneto}, there is a crucial need to fully categorise all methods of magnetic control for these materials. Additionally, Mn$_2$Au provides a generous platform for both high and low level model development and experimental research: its high N\'eel temperature, simple collinear magnetism, and large magnetic moment \cite{Barthem2013} make it an engaging target for applications. 

Recent work \cite{Freimuth2021} has presented a third option for direct N\'eel vector control of AFMs: the induction of staggered fields using direct laser excitation. The experimental evidence of these  torques has been seen in ferromagnetic (FM) materials \cite{N_mec_2012,Choi2017}. Only recently has the same formalism \cite{Freimuth2016} been applied to AFMs, specifically Mn$_2$Au \cite{Freimuth2021,Merte2023}. There, the induced fields are generated from the inverse Faraday Effect (IFE) \cite{Freimuth2021}. The frequency dependence of the induced staggered magnetic fields is calculated for optical and THz wavelengths, and is shown to generate a net non-staggered torque. This torque could potentially switch the AFM order parameter, an idea we consider in detail in the present article. Here we present atomistic spin dynamics simulations of an optical frequency excitation from ultrafast laser pulses on Mn$_2$Au using the coupling scheme suggested in Freimuth \textit{et al} \cite{Freimuth2021}. To distinguish between the other optical torque methods through spin transfer techniques, we call this generated torque a laser optical torque (LOT). Unlike THz pulses which are shown to induce both LOT and SOT fields, the optical frequency is too far above the AFM frequency to excite SOT dynamics. Thus, we demonstrate the possibility to switch the N\'eel vector in AFMs using purely LOTs. Additionally, we provide a method using the novel symmetry relationships in the LOT to preferentially control the switching direction of the N\'eel vector, allowing for deterministic, non-toggle all-optical siwtching (AOS) in AFMs.

\section{Methodology}

\begin{figure}[t]
  \includegraphics[width=\columnwidth]{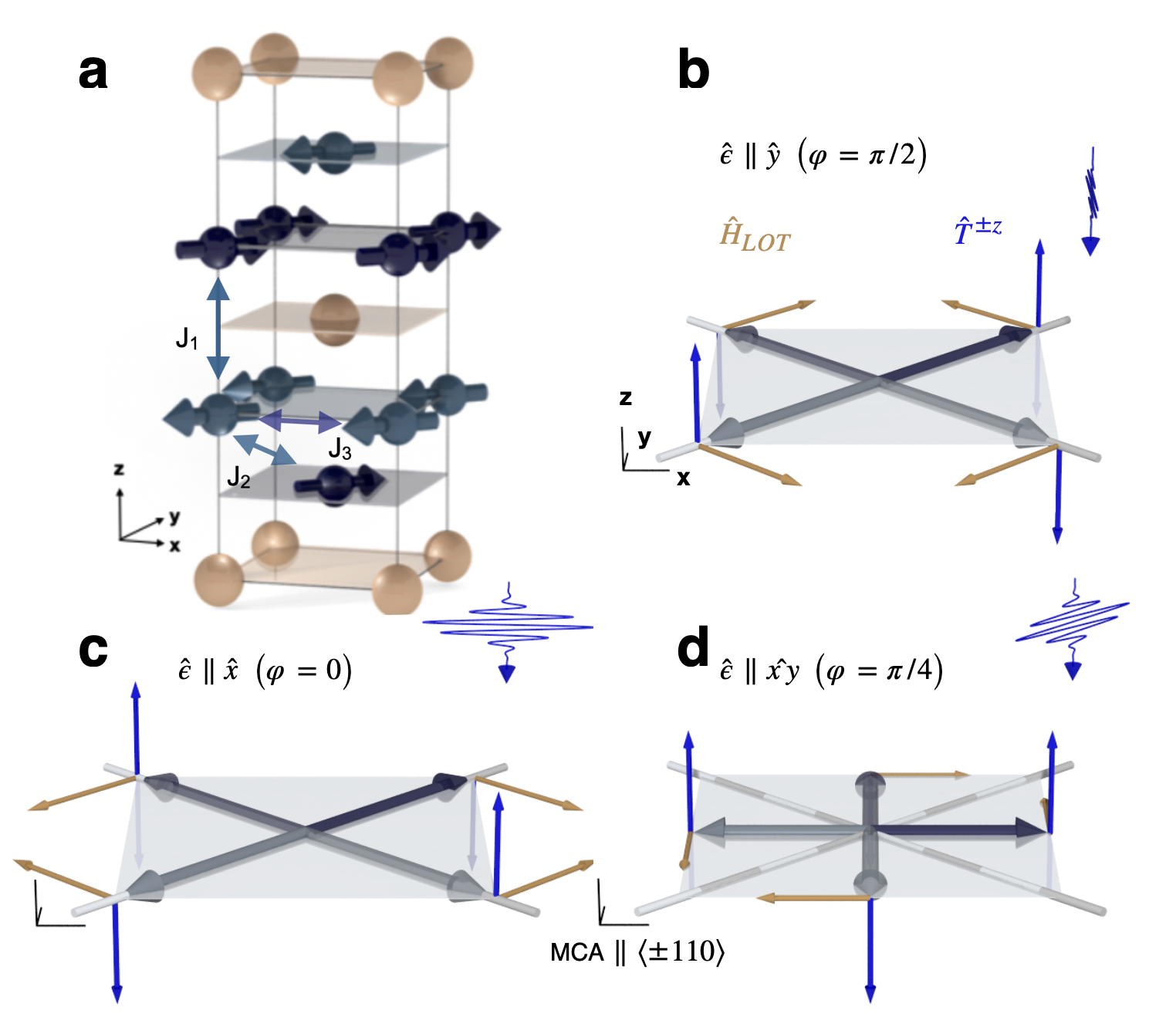}
    \caption{(a): Unit cell with exchange interactions. (b-d): Diagram of net torques from Eq. \ref{eq:lot} with linearly polarised light parallel to the azimuthal angle $\varphi$ from $x$ and fourth order in-plane magneto-crystalline anisotropy (MCA) along $xy$. (Purple): magnetisation vector, (gold): induced field, (blue): resultant torque. (b): electric field polarised along $y$ axis. (c): electric field polarised along $x$ axis (same tensor as in (b) but with opposite symmetry). (d): electric field polarised along $\langle110\rangle$. (b) and (c) produce the maximal torque for switching. For consistency we use the polarisation in (b) in our simulations.}
    \label{fig:tensors}
\end{figure}

We perform atomistic spin dynamics simulations in Mn$_2$Au based on the Landau-Lifshitz-Gilbert (LLG) equation as implemented in the code VAMPIRE \cite{Evans_2014}. Fig.\ref{fig:tensors} illustrates Mn$_2$Au unit cell implemented in the atomistic simulations. The effective Heisenberg spin Hamiltonian includes the ferromagnetic (FM) and antiferromagnetic (AFM) exchange interaction terms, two-ion anisotropy mediated by the Au sublayers, fourth-order out-of-plane anisotropy, and the fourth-order rotational in-plane anisotropy given by 

\begin{equation}
\label{eq:spin-hamiltonian}
\begin{split}
& \mathcal{H}= -\frac{1}{2}\sum_{ij} \mathbf{S}_i \boldsymbol{J}_{ij} \mathbf{S}_j -k_{4\perp} \sum_i \left(S_{i,z}^4 - \frac{30}{35} S_{i,z}^2\right) \\
&-k_{4rot} \sum_i \sin^4 \theta \cos 4\phi - \mu_s \sum_i \mathbf{S}_i \cdot \mathbf{H}_{LOT,i}.
\end{split}
\end{equation}

The local moment directions are given by unit vectors $\mathbf{S}_i$ with local moment length $\mu_s$. $\theta$ gives the polar angle of magnetisation, and $\phi$ the rotational angle of magnetisation from the $x$-coordinate. The magneto-crystalline anisotropies (MCAs) are implemented in Cartesian coordinates using the spherical harmonic formulation given in ~\cite{Collings_2023} for accurate four-fold rotational symmetry. Also included is the Au mediated two-ion anisotropy terms, creating a vectorial form of the exchange constants such that $\boldsymbol{J}_{ij} = \langle\boldsymbol{J}_{xx}, \boldsymbol{J}_{yy}, \boldsymbol{J}_{zz} + k_2^\perp\rangle$ \cite{Shick_2010} for exchange interactions across Au sites (footnote \footnote{Simulations of the two-ion anisotropy included as a traditional out-of-plane uniaxial anisotropy show no discernible differences in the phase diagrams, but the two-ion anisotropy is known to have a different temperature scaling; the effects of this at finite temperature will be studied in future}). Table I summarizes the parameters used in the simulations. Shick \textit{et al}~\cite{Shick_2010} discussed situations where thin films of Mn$_2$Au have an additional in-plane strain anisotropy, creating a preferential 180 degree orientation along $\langle 100 \rangle$ or $\langle 010 \rangle$. While SOTs/STTs have been used to switch 180 degree domains in Mn$_2$Au \cite{Roy2016,Weienhofer2023}, we confine our simulations to 90 degree domains.

\begin{table}[!tb]
\centering 
\label{table:constants}
\begin{ruledtabular}
\begin{tabular}{c c c c c }
Interactions & $J_{xx}$ & $J_{yy}$ & $J_{zz}$ & Unit \\
\hline 
$J_1$ & -1.46923 & -1.46923 & -1.45932 & $10^{-20}$ J/link \\
$J_2$ & -1.09430 & -1.09430 & -1.08691 & $10^{-20}$ J/link \\
$J_3$ &  0.31826 &  0.31826 &  0.31826 & $10^{-20}$ J/link \\
\hline 
 & Parameter & \multicolumn{2}{c}{Value} & Unit \\
\hline 
&$\mu_s$         & \multicolumn{2}{c}{3.72}                     & $\mu_B$ \\
&$k_{2\perp}$     & \multicolumn{2}{c}{$-1.303 \times 10^{-22}$}  & J/atom \\
&$k_{4\perp}$           & \multicolumn{2}{c}{$3.71 \times 10^{-25}$}     & J/atom \\
&$k_{4rot}$        & \multicolumn{2}{c}{$1.855 \times 10^{-25}$}     & J/atom \\
&$T_N$        & \multicolumn{2}{c}{$1220$}     & K\\
\end{tabular}
\end{ruledtabular}
\caption{Exchange and anisotropy constants from Khmelevskyi \cite{Khmelevskyi_2008} and Shick \cite{Shick_2010}, in line with experimental values calculated in \cite{Barthem2013}.
}
\end{table}
 
Optically-induced torques show strong crystal symmetry and frequency dependent coupling to the polarised electric field components of the laser. A full analysis of the symmetry requirements in the Mn$_2$Au bulk crystal was previously presented by Freimuth \textit{et al} \cite{Freimuth2021} based on the Keldysh non-equilibrium  formalism. There, a linearly or circularly polarised laser pulse was shown to be capable of inducing a torque acting on the N\'eel vector parameter $\mathcal{\bm{L}}$ via staggered magnetic fields induced by the inverse Faraday effect, which couple to the Mn spins in the distinct sublattices of the AFM. The magnitude and spatial symmetry of the observed torque depends both on the local orientation of the N\'eel vector $\mathcal{\bm{L}}$, as well as the electric field $\bm{\epsilon}$ direction of the applied optical pulse. To drive magnetic switching using an in-plane torque, the induced field must be at least equal to the in-plane rotational anisotropy field value of $10.3$ mT \cite{Roy2016}. Assuming a constant linear relationship between torque and intensity, this would require pulses on the order of 2000 mJ/cm$^2$. Thus, this work seeks to apply out-of-plane torques to take advantage of the exchange enhancement characteristic of antiferromagnetic switching \cite{Roy2016,Gomonay_2010}. 

The  torque tensors \cite{Freimuth2021}  
depend on the vector components of the electric polarisation and AFM order parameter:
\begin{equation}
\label{eq:torque-sum}
    T_i = \frac{a_0^3 I}{c}\left(\frac{\mathcal{E}_H}{\hbar  \omega} \right)^2 \text{Im}\sum_{jklm} \chi_{ijklm} \epsilon_j \epsilon_k^* \mathcal{L}_l \mathcal{L}_m 
\end{equation}
where $\epsilon_0$ is the vacuum permittivity,  $\hbar$ the reduced Plank's constant, $m$ the mass of the free electron, Bohr radius $a_0 = 4\pi \epsilon_0 \hbar^2/(me^2)$, $c$ is the speed of light, $e$ the fundamental electron charge, $I = \epsilon_0 c E_0^2/2$ is the laser intensity calculated from the electric field component $E_0$, $\mathcal{E}_H$ the Hartree energy, $\epsilon_j$ is the $j$th Cartesian component of the electric field, and $\mathcal{L}_j$ is the $j$th Cartesian component of the N\'eel vector and $\chi_{ijklm}$ are the corresponding susceptibility components. Importantly, the induced torque sums as the square of the vector components, so the resultant symmetry of the torque can be non-trivial. 
The Cartesian vector components of $\mathbf{\epsilon}, \mathbf{\mathcal{L}} = (\text{sin}\theta \text{cos}\phi, \text{sin}\theta \text{sin}\phi, \text{cos}\theta)^T$ (we will use $\varphi$ as the azimuthal angle from $x$ for the laser polarisation and the azimuthal angle of the N\'eel vector as $\phi$).
Freimuth \textit{et al.} \cite{Freimuth2021} present thirty tensors which are allowed by the Mn$_2$Au orbital symmetry and which produce a torque perpendicular to the N\'eel vector. Since  Mn$_2$Au has in-plane magnetisation, we disregard tensors corresponding to an out-of-plane N\'eel vector component to good approximation; likewise, since we are interested in exchange enhanced precessional switching, we choose the geometry with the generated torque to be out-of-plane. Lastly, Freimuth \textit{et al} find that light incident normal to the AFM plane with electric field linearly polarised along the in-plane angle $\varphi$ produces the largest torque values. Our chosen laser geometry ($\varphi = \pi/2$, Fig. \ref{fig:tensors}b) thus requires two susceptibility tensors to fully describe the excited torque (tensors 4 and 24 in Freimuth \textit{et al} \cite{Freimuth2021}), which can be approximated to follow the trigonometric relation $\text{sin}(2\varphi - 2\phi)$.

Taking the physical and experimental constants into the variable $\tau(I,\omega)$, the sum of Eq.~\ref{eq:torque-sum} can be written as 
\begin{equation}
    \label{eq:lot}
    \mathbf{H}_{\mathrm{LOT}} = \frac{1}{\mu_s}\tau(I,\omega) \sin(2\varphi-2\phi)\hat{z}\times \mathbf{S}
\end{equation}

The induced fields from various light polarisation angles are illustrated in Fig. \ref{fig:tensors}. The staggered fields then lead to a non-staggered effective torque. Fig.~\ref{fig:tensors}b and c show the polarisation with the maximal torque useful for switching. Fig.~\ref{fig:tensors}d shows the light polarisation along $\langle 110 \rangle$ in \cite{Freimuth2021}, inducing an out-of-plane torque relationship which does not produce a torque useful for switching, as the maximal torque generated is when the N\'eel vector is along the hard axis. Thus, in the following simulations we use linearly polarised light with polarization parallel to $\langle010\rangle$ (Fig. \ref{fig:tensors}b).

To simulate the influence of the LOT generated by an ultrafast laser pulse, we scale the laser intensity following a Gaussian time-dependent profile, with the pulse duration giving the time at full-width at half-height. Eq. \ref{eq:torque-sum} provides a linear relationship with laser intensity, so to model the time dependence of an ultrafast laser pulse we vary the intensity with a Gaussian profile. A linearly polarised pulse with $\bm{\epsilon}|| \langle 010 \rangle$ and intensity $I=10$ GW/cm$^2$, set at a photon energy of $h\nu=1.55$ eV, will produce a LOT of magnitude $\approx 12 \times 10^{-24}$ J (Eq. \ref{eq:lot}), which corresponds to an effective field of 17.3 mT on each magnetic moment, canting the local N\'eel vector out-of-plane. For comparison, we can also consider a sub-optimal switching polarisation geometry with the laser polarisation rotated further along the polar angle: an electric field polarisation along $xz$, frequency of $1.5$ eV, intensity of 10 GW/cm$^2$, and N\'eel vector along $x$ produces an \textit{in-plane} torque on each magnetic site of $\approx 10^{-24}$ J $(H_{LOT} = 0.05$ mT) \cite{Freimuth2021}. 

\section{Results \label{sec:results}}

\begin{figure}
    \centering
    \includegraphics[width=\columnwidth]{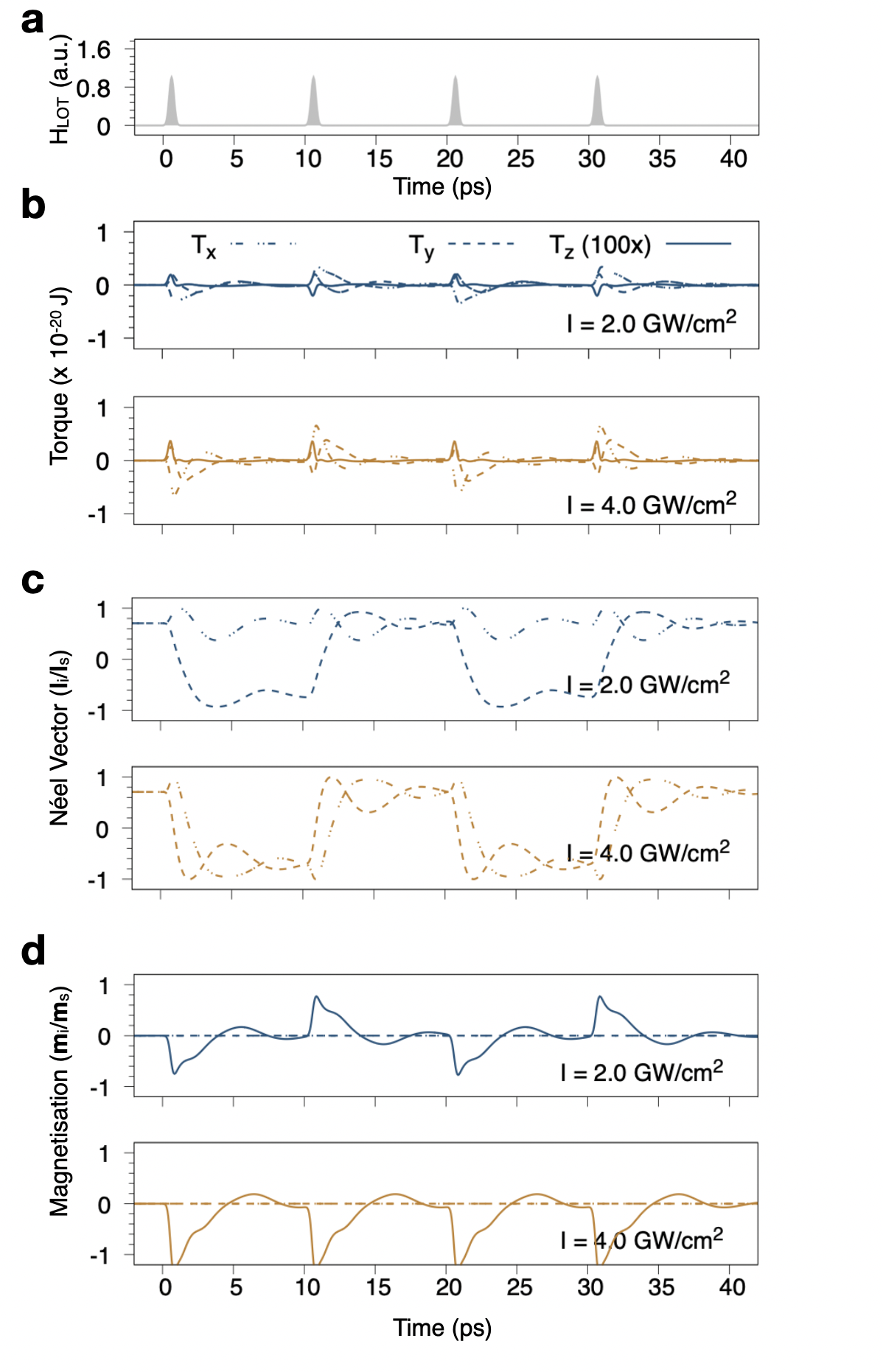}
    \caption{Toggle switching of the N\'eel vector with multiple optical pulses (a): normalised field intensity scaled to laser pulse (400 fs) 8 ps apart. (b): total exchange torque. Note that the z torque has been scaled 100x for visibility. (solid): $T_x$; (dashed): $T_y$; (dash-dot): $T_z$. (c): $x$ and $y$ N\'eel vector components $\boldsymbol{L}_i = (\boldsymbol{m}_1-\boldsymbol{m}_2)/2$. (d): Net magnetisation $\boldsymbol{m}=(\boldsymbol{m}_1+\boldsymbol{m}_2)/2$ (scaled for visibility). Blue lines show 90 degree toggle switching for four sequential pulses. Gold shows 180 degree toggle switching for four sequential pulses. The sign of the induced $z$ net magnetisation could be read experimentally to tell the direction of rotation.
    }
    \label{fig:sot-toggle}
\end{figure}

\begin{figure}
\includegraphics[width=\columnwidth]{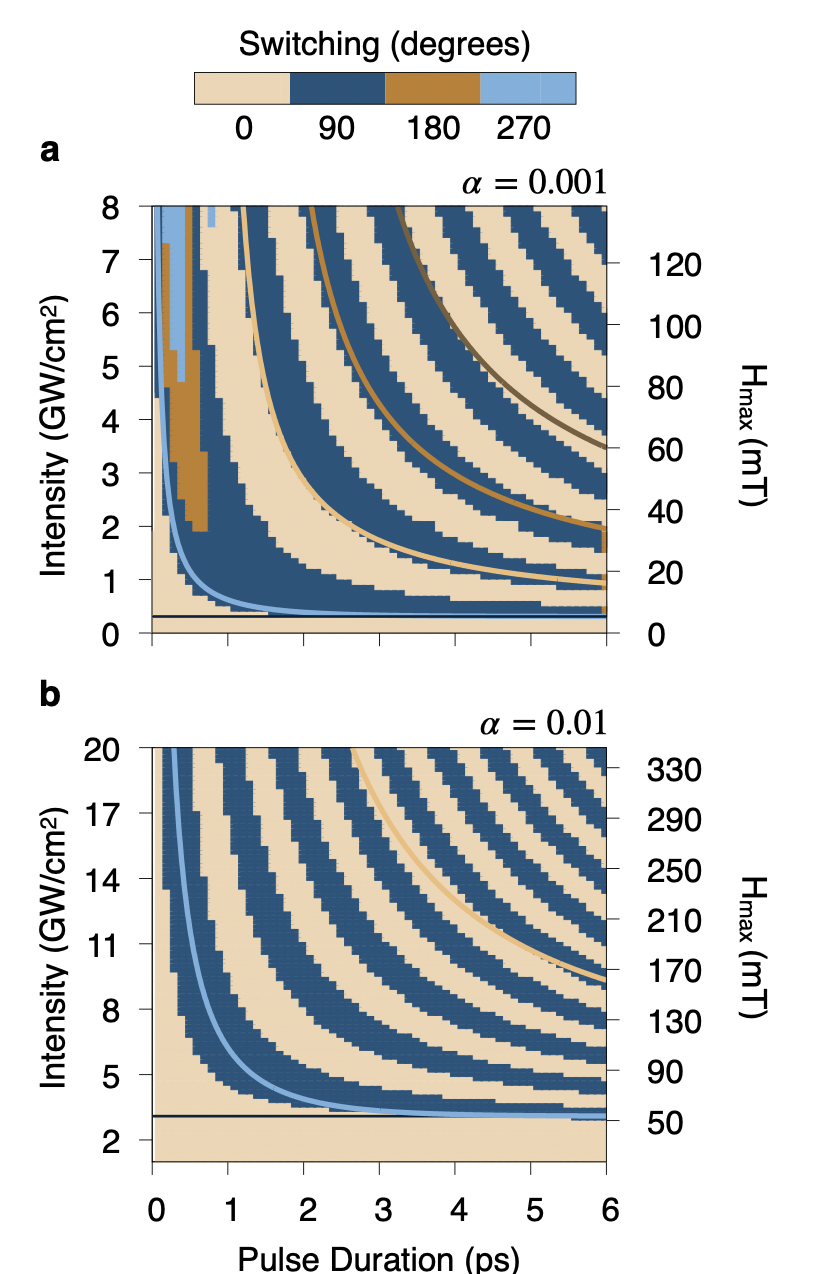}
    \caption{Switching phase diagram in terms of laser intensity and duration. Colours represent the end change of angle from $\langle110\rangle$. The corresponding maximum field strength for the laser intensity is given on the second y-axis. Solid colour lines show the analytic $H_{crit}$ values from Eq. \eqref{eq:coth-n} (a) for $\alpha = 0.001$. Black line: $H_{crit} = \omega_{4\parallel}/(2\gamma)=5.16$ mT. (b) Black line: $H_{crit}$ with 10x factor from $\alpha = 0.01$.}
    \label{fig:Jenkins-sr-0.001-pd}
\end{figure}

The LOT governed by Eq.~\eqref{eq:lot} produces precessional switching of the N\'eel vector in Mn$_2$Au (Fig. \ref{fig:sot-toggle}). The LOT precessional switching follows the well-characterised \textit{exchange-enhancement} generated by the out-of-plane canting as seen in SOT switching, \cite{Roy2016}. Fig. \ref{fig:Jenkins-sr-0.001-pd} presents the phase diagrams as a function of laser intensity and duration. The colour variation shows intervals of 90 degree switching, typical for precessional switching. The diagrams are presented for two damping parameters $\alpha=0.001$ and $\alpha=0.01$. Importantly, the switching can be produced by ultrafast laser pulses of several picosecond duration and even below. The absorbed fluence of the sample is approximately linear with intensity and pulse duration \cite{Freimuth2016,Merte2023}. Remarkably, we observe an absorbed fluence of 0.5 mJ/cm$^2$ ($ I=1 \; \text{GW/cm}^2,\;H_{crit} = 6.34 \; \text{mT} $) is sufficient to induce switching on the sub-picosecond timescale for the smaller damping value of $\alpha=0.001$. Increasing the damping parameter shows a linear dependence in the critical field (see Fig. \ref{fig:sot-toggle}b), commensurate with the theory of SOT switching \cite{RamaEiroa2022}. Also remarkable is the induced magnitude and sign of the non-zero net magnetisation seen during the switching process (Fig. \ref{fig:sot-toggle} below), which could be externally detected. 

\begin{figure}
    \centering
    \includegraphics[width=0.4\textwidth]{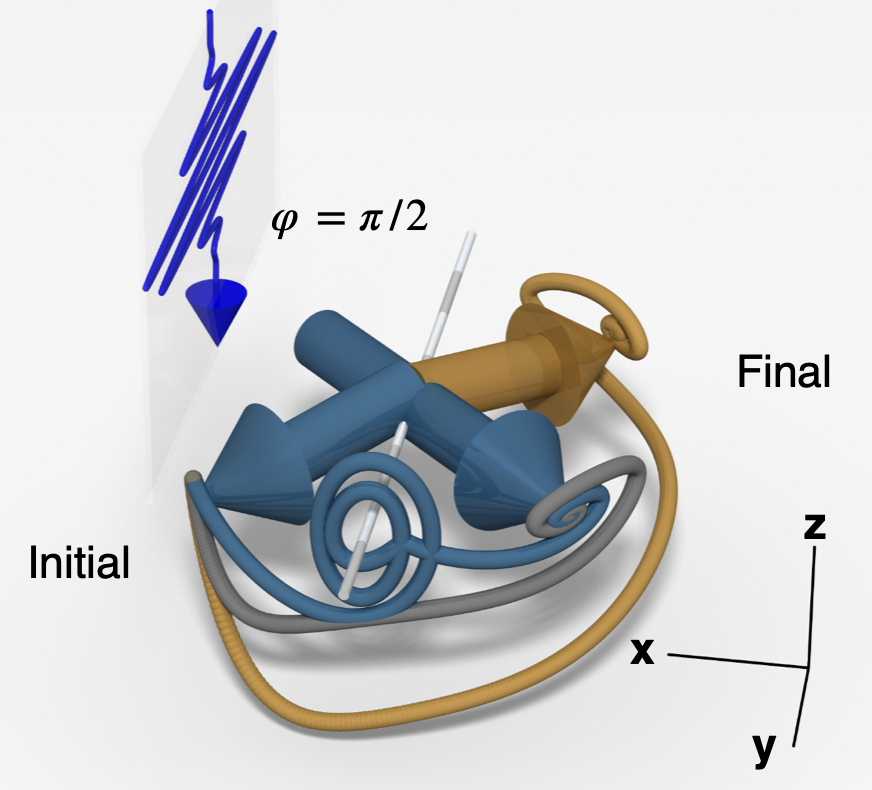}
    \caption{Simulation trace of sublattice magnetisation vectors switching from $\langle110\rangle$ to $\langle-110\rangle$ with $\boldsymbol{\epsilon}\parallel\langle 010\rangle$. (Bar): precession axis. (gold): $t_p=$ 400 fs, $I=2$ GW/cm$^2$, (grey): $t_p = 400$ fs, $I= 4$ GW/cm$^2$, and (blue): $t_p=$ 3 ps, $I=5$ GW/cm$^2$ ($z$ magnetisation scaled for visibility).}
    \label{fig:spin-track}
\end{figure}

Analytically, the dynamic solution for the action of collinear antiferromagnets under staggered in-plane fields is discussed in  \cite{Roy2016}. Here, the critical field ($H_{crit}$) for switching can be evaluated as a combination of the characteristic exchange, anisotropy, and induced magnetic fields. Following the method of \cite{Roy2016} - modified for the LOT induced field - the dynamics of the N\'eel vector in the $xy$ plane can be described by the following equation:

\begin{equation}
\label{eq:macrospin-ode}
\ddot{\phi}+\frac{\omega_R^2}{4}\text{cos}4\phi-\gamma \omega_e \frac{\tau(I,\omega)}{\mu_s} \text{sin}(2\varphi-2\phi) + 2\alpha\omega_e \dot{\phi} = 0 
\end{equation}
where $\omega_e = |5J|\gamma/\mu_s$ is the AFM exchange frequency, $\omega_{4\parallel} = 2\gamma k_{4rot}$ is the fourth order in plane anisotropy frequency, $\alpha$ is the atomistic damping parameter, and $\omega_R = \sqrt{2\omega_e \omega_{4\parallel}}$. The parameter $\tau_z/\mu_s$ corresponds to the amplitude of the LOT field. If the N\'eel vector is only considered in the interval of its azimuthal angle $\phi \in [0,\pi/4]$, the critical field for infinitely long pulse lengths is $H_{crit}=\omega_{4 \parallel}/(2\gamma)$: a factor of 2 larger than for SOT \cite{Roy2016,RamaEiroa2022}, due to the $\text{sin}2\phi$ dependence of the LOT field. Analytically, our constants predict an $H_{crit}=5.16 \; \text{mT}$, matching well with our simulations (see horizontal line in Fig. \ref{fig:sot-toggle}a) . 

For short pulse lengths the critical switching field can be evaluated as 
\begin{equation}
    \label{eq:coth-single}
    \frac{H_{crit}}{H_{crit}(\tau \rightarrow \infty)}= \text{coth}\left(\frac{2t_p}{\tau_e\sqrt{2\pi}}\frac{\omega_R}{\omega_e}\right),
\end{equation}
where $\tau_e \approx \pi/(2 \omega_e)$ is the timescale of the exchange interaction to precess the N\'eel vector 90-degrees, which is then scaled in Eq. \ref{eq:coth-single} by $\sqrt{2\pi}/2$ to account for the Gaussian profile of the laser pulse. Then, $\tau_p$ can be called the characteristic pulse duration $\tau_p = 2\tau_e \omega_e/\omega_R$ such that $\text{coth} (2) \approx 1$. 

Unlike the SOT considered in \cite{Roy2016,RamaEiroa2022}, the LOT has the additional feature of changing sign during the switching: the intrinsic spatial symmetry defined in equation Eq. \eqref{eq:lot} ensures the induced LOT changes its sign for any 90-degree rotation of the N\'eel vector. This allows for both clockwise and counter-clockwise switching by means of the same laser polarisation. In contrast is the SOT torque, where the direction of the applied current needs to be constantly inverted \cite{Bodnar2018} to change the handedness of rotation. Additionally, prolonged laser pulses will force precession of the magnetisation around the vector $\parallel \hat{\varphi}$, rather than continuously drive the order parameter in plane. Only for high intensities and short pulse duration is the inertia generated by the exchange torque susceptible to overshooting, either to the $\langle\overline{11}0\rangle$ (180 degrees) or even $\langle 1\overline{1}0 \rangle$ (270 degrees) states (orange and light blue color in Fig. \ref{fig:Jenkins-sr-0.001-pd}). This is visible in the switching diagram for $\alpha=0.001$, and illustrated by the gold track in Fig. \ref{fig:spin-track}. This effect disappears for pulse lengths beyond 1 ps: rather than continually drive the precessional switching using exchange enhancement, the long pulse duration has the N\'eel vector precessing along the laser polarisation axis (blue curve in Fig. \ref{fig:spin-track}). The timing of these precessions follows the characteristic exchange period, and can be predicted by modifying Eq. \ref{eq:coth-single} to take into account the additional precessions around the LOT polarisation vector: for $n$ precessions, the minimum pulse time to switch the N\'eel vector at high intensity is proportional to the exchange precession timescale and precession number, $4n\tau_e\sqrt{2\pi}$. This gives a rough approximation to the critical field and laser pulse time:

\begin{equation}
\label{eq:coth-n}
    H_{crit}(\tau, n) = nH_{crit}^{\tau \rightarrow \infty}\text{coth}\left(\frac{2 t_p \omega_R}{n\tau_e \sqrt{2\pi}}-\frac{4n\tau_e \sqrt{2\pi}}{t_p}\right)
\end{equation}

The analytical estimates of critical fields for switching are presented in Fig. \ref{fig:Jenkins-sr-0.001-pd}a, showing a good agreement for $n \leq 3$ with the lowest damping value, especially considering the approximations made. Very importantly, the torque's $\text{sin}(2\phi)$ dependence on the magnetisation direction allows sequential laser pulses to reverse the switching direction. Fig. \ref{fig:sot-toggle} illustrates our results with a sequence of laser pulses showing a toggle switching of the N\'eel vector, which is reversed with each laser pulse application. This is in direct contrast to traditional STT \cite{Huang2022} and SOT \cite{Roy2016} experiments, where sequential torques continuously drive the N\'eel vector either clock-wise or counter-clockwise.

In the Supplementary Information we present the estimations of the switching diagram for a different set of atomistic parameters of Mn$_2$Au used in \cite{Selzer2022}. While the exchange parameters are similar, the in-plane cubic anisotropy in \cite{Selzer2022} is two orders of magnitude larger than our case \cite{Shick_2010,Barthem2013}. The overall effect is that approximately 10 times larger laser intensities would be needed for LOT switching and the validity of Eq. \eqref{eq:coth-single} is further reduced. Finally, Ref. \cite{Selzer2022} uses RKKY-type Hamiltonian interactions for including exchange interactions with more than nearest neighbors (NN) up distances of $8.91 \;\AA$. This has the effect of increasing the total AFM exchange by $30 \% $ and the FM exchange by $17 \%$. To study the effect of long-range interactions on the switching, we extend our atomistic Hamiltonian up to $5th$ NN interactions for both Hamiltonians. This appears to have no substantial effect on the phase diagram or switching time, despite decreasing the characteristic period of the exchange interaction $\tau_e$. 

\begin{figure}
    \centering
    \includegraphics[width=\columnwidth]{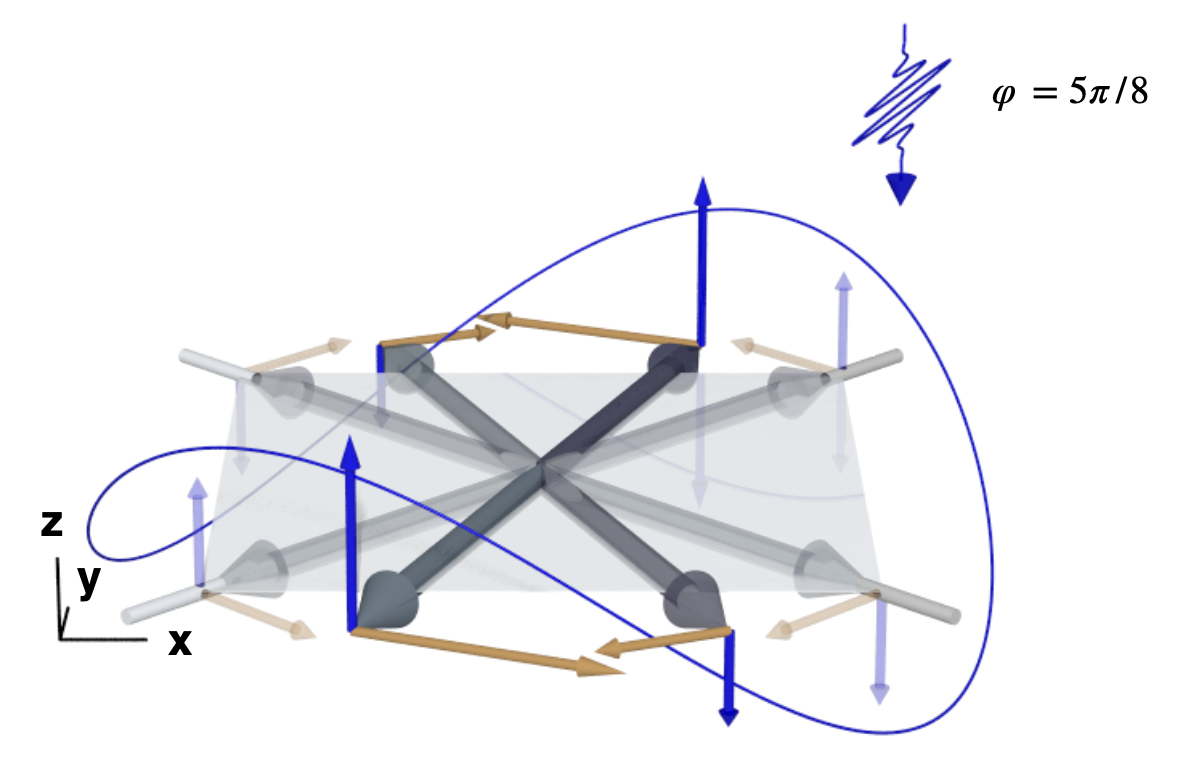}
    \caption{Torque and field diagram for electric field polarisation along $5\pi/8$. (blue line): relative size and sign of the out of plane torque as a function of the magnetisation direction. Opaque shows comparable torque and field sizes further along the path of motion from the easy axis. Note the unequal torque generated between clockwise and counter-clockwise motion.}
    \label{fig:det-torque}
\end{figure}

\begin{figure}
    \centering
    \includegraphics[width=\columnwidth]{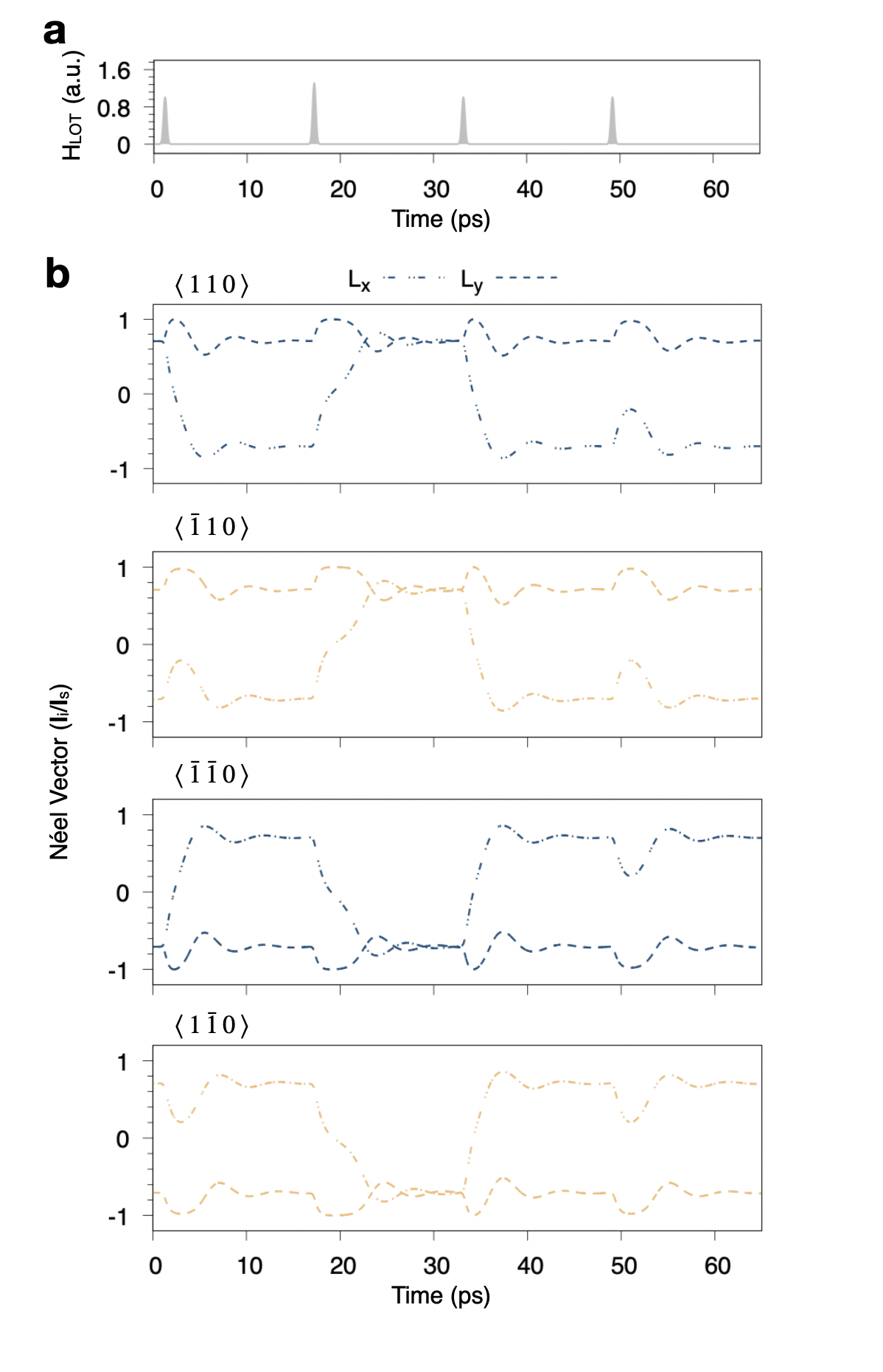}
    \caption{Dynamics of the N\'eel order parameter for four sequential pulses 16 ps apart. Each panel shows a the N\'eel vector starting along each easy axis. The second pulse has an intensity of 2.6 GW/cm$^2$; all others are 2 GW/cm$^2$. Light polarisation is parallel to the angle $\varphi = 5\pi/8$. 
    }
    \label{fig:determ-switch}
\end{figure}

We have shown above that the quadratic reliance on the electric field polarisation vector of the generated torque in Eq. \ref{eq:lot} allows for rotation of the laser vector to shift the maximum torque away from the easy axis. Contrary to the toggle switching caused by $\boldsymbol{\epsilon} \parallel \langle100\rangle$ or $\langle 010\rangle$ displayed in Fig. \ref{fig:sot-toggle}, shifting the azimuthal angle of the laser polarisation will create an asymmetric torque profile for the path of motion from the four easy axes. Instead of SOT and STT methods which induce symmetric torques from each easy axis--rotating the N\'eel vector uniformly clockwise or counter-clockwise--the generated torque is asymmetric along the path of motion of the magnetisation vector. Thus, the magnetisation will experience a local torque maxima when starting from only two of the four easy axis directions, giving a preference between clockwise and counter-clockwise switching (Fig. \ref{fig:determ-switch}). This breaks the four-fold degenerate easy axis into "large" and "small" generated torques. 

The anisotropic magneto-resistance (AMR),  typically used to determine the magnetic state in antiferromagnets\cite{Bodnar2018}, depends only on the parallel component of the N\'eel vector, condensing the four-fold symmetry of the MCA into two states "aligned" and "anti-aligned" with the resistance measurement vector. Thus, the $\pi/4$ and $-\pi/4$ N\'eel states can be considered equivalent (as with $3\pi/4$ and $-3\pi/4$). This equivalence, combined with the asymmetric torque generated by a laser polarisation shifted away from $\langle 010\rangle$ or $\langle 100\rangle$, allows the possibility of deterministic N\'eel vector control by using sequential pulses of varying pulse times or intensities to preferentially switch the AFM order parameter between \textit{aligned} and \textit{anti-aligned} states with the (AMR) measurement.

Fig. \ref{fig:determ-switch} details the N\'eel vector dynamics following sequential laser pulses with starting AFM order parameters along each of the MCA easy axis with laser polarisation parallel to the $5\pi/8$ azimuthal angle. 16 ps is given to allow the magnetisation to return to equilibrium, with the second pulse being larger than the others. Fig. \ref{fig:determ-switch}(c) and (e) fail to switch with the first, "small" pulse, but switch with the second "large" pulse. Now that the magnetisation has been reoriented to the quadrant which experiences the maximal torque, it does successfully switch following a small pulse. This dependence is shown with Fig. \ref{fig:determ-switch}(b) and (d) as well: since they start their magnetisation in the maximal torque quadrant, they switch following the first "small" pulse, as well as the second and third pulses, but not the fourth, since the magnetisation has moved to a minimum torque quadrant.

\begin{figure}
    \centering
    \includegraphics[width=\columnwidth]{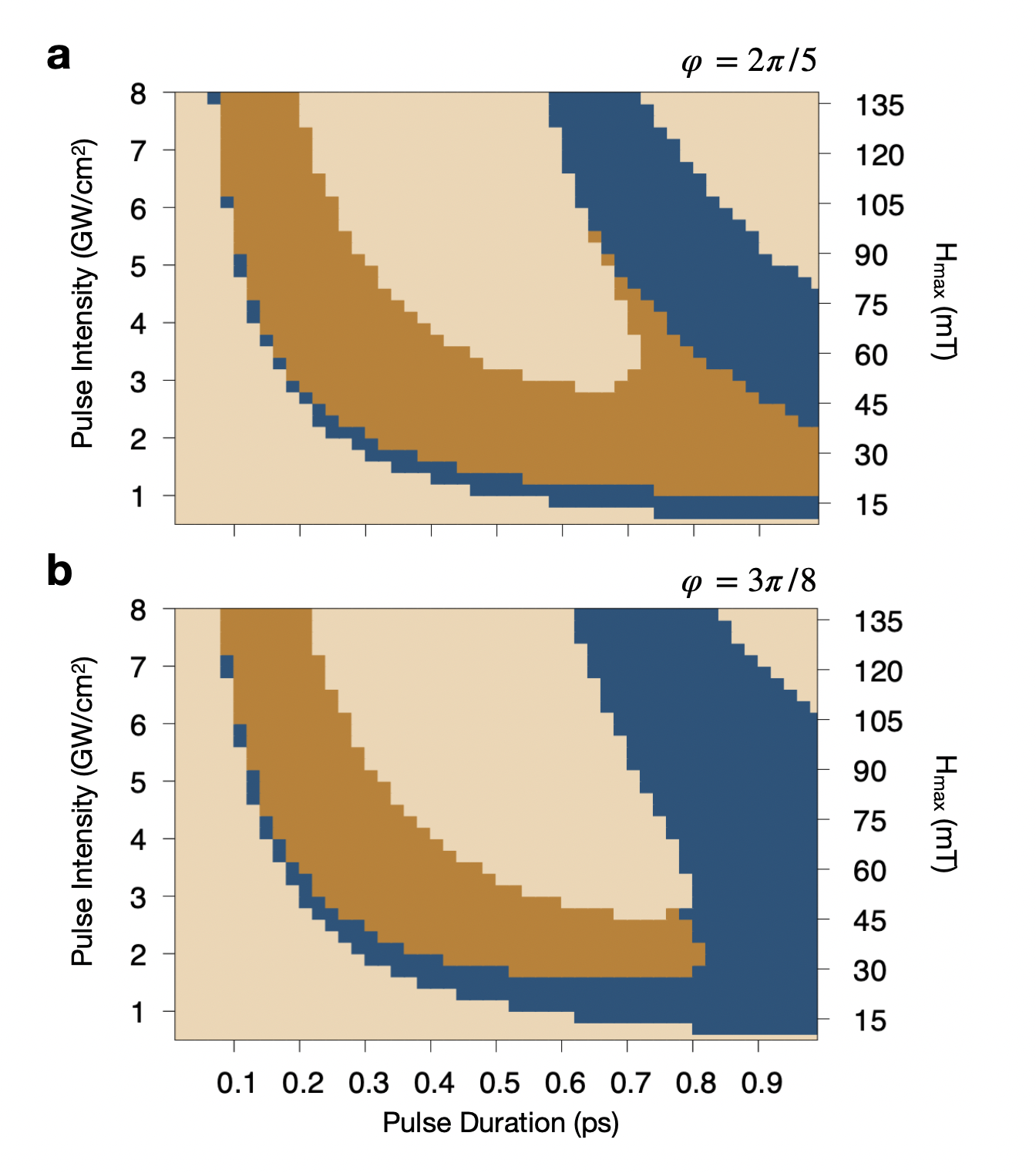}
    \caption{Switching phase diagram for variable pulse intensity and duration with laser polarisation along $2\pi/5$ and $3\pi/8$. (tan): no switching. (dark blue): single pulse produces counter-clockwise switching only. (gold): single pulse toggle switching. $\alpha = 0.001$.}
      \label{fig:det-diagram}
\end{figure}

Fig. \ref{fig:det-diagram} shows the phase diagram of laser pulse intensity and duration which distinguishes between non-toggle, preferential switching and toggle switching. Fig. \ref{fig:det-diagram} has $\varphi=5\pi/8$ (opposite symmetry to Fig. \ref{fig:det-torque}) and $\varphi=3\pi/8$.

\section{Discussion and Conclusion}

By means of atomistic spin dynamics simulations with LOT we predict the possibility for AFM switching on the sub- and low picosecond timescale. The unique symmetry relations of this torque prioritise 90-degree switching for long pulse times, but also allow 180 or 270-degree switching provided sufficiently large pulse intensities are used. Also of interest is the reversible nature of the switching process: repeated laser pulses do not drive the order parameter continuously clockwise (counter-clockwise) like SOT/STT excitations. Rotation of the laser polarisation to generate a quadrant-asymmetric torque introduces an additional level of control to the switching process, allowing for preferential, non-toggle switching. These results suggest the significant opportunity the LOT could provide for deterministic AOS in AFM spintronics.

The efficiency of the LOT may be preferred to other AOS methods. The pulse intensity and fluence necessary to switch on the sub-picosecond timescale using the LOT is 1 GW/cm$^2$ and 0.5 mJ/cm$^2$, respecitviely. This is compared to the fluence for GdFeCo AOS in \cite{Radu2011} of 4.4 mJ/cm$^2$ and the 6.51 mJ/cm$^2$ for Fe ultrafast demagnetisation generating spin current which is used for the STT switching in Ref. \cite{Weienhofer2023}. Additionally, the LOT has the advantage of not needing to be applied over fs time scales to generate the spin current injection \cite{Weienhofer2023,Huang2022}; indeed, the critical field necessary for switching can be reached on small picosecond timescales with a fluence as low as 0.65 mJ/cm$^2$. 

Lastly, we extend the comment in Freimuth \textit{et al} that this LOT is not unique to Mn$_2$Au \cite{Freimuth2021}. The crystal and orbital symmetry rules of the antiferromagnet play the largest role in determining the coupling between the laser excitation and the spin-orbital interaction. As the study and characterisation of altermagnetism continues, more materials of relevant symmetry should be included in the theoretical and experimental study of optically generated spin torques. 

\begin{acknowledgments}
The authors acknowledge funding from the European Union’s Horizon 2020 research and innovation programme under the Marie Skłodowska-Curie International Training Network COMRAD (grant agreement No 861300). The atomistic simulations were undertaken on the VIKING cluster, which is a high performance compute facility provided by the University of York. F.F. and Y.M. acknowledge the funding by the Deutsche Forschungsgemeinschaft (DFG, German Research Foundation) $-$ TRR 173/2 $-$ 268565370 (project A11) and Sino-German research project DISTOMAT (MO 1731/10-1) of the DFG. The work of O.C.-F. has been supported by DFG  via CRC/TRR 227, project ID 328545488 (Project MF).

\end{acknowledgments}

\bibliography{bib1}

\end{document}